# Enhancement of Thermally Injected Spin Current through an Antiferromagnetic Insulator

Weiwei Lin,[1,*] Kai Chen,[2] Shufeng Zhang[2] and C. L. Chien[1,†]

[1]*Department of Physics and Astronomy, Johns Hopkins University, Baltimore, Maryland 21218, USA.*

[2]*Department of Physics, University of Arizona, Tucson, Arizona 85721, USA*

**Abstract**

We report large enhancement of thermally injected spin current in normal metal (NM)/antiferromagnet(AF)/yttrium iron garnet(YIG), where a thin AF insulating layer of NiO or CoO can enhance spin current from YIG to a NM by up to a factor of 10. The spin current enhancement in NM/AF/YIG, with a pronounced maximum near the Néel temperature of the thin AF layer, has been found to scale linearly with the spin-mixing conductance at the NM/YIG interface for NM = *3d*, *4d*, and *5d* metals. Calculations of spin current enhancement and spin mixing conductance are qualitatively consistent with the experimental results.

[*]wlin@jhu.edu

[†]clchien@jhu.edu

Pure spin current phenomena and devices are new advents in spin electronics [1,2]. A pure spin current has the unique attribute of delivering spin angular momentum without a net charge current thus with higher energy efficiency. A pure spin current can be generated by several mechanisms, including spin Hall effect [1-3], lateral spin valves [4,5], spin pumping [6,7], and longitudinal spin Seebeck effect (LSSE) [8,9]. The inverse spin Hall effect (ISHE) in a metal can detect a pure spin current by converting it into a charge current with resultant charge accumulation [3,10]. Inevitably a spin current decays as it traverses through a material on the scale of the spin diffusion length $\lambda_{SF}$, which depends on the strength of the intrinsic spin orbit interaction and the quality of the material [5]. The transmission of a spin current across an interface between two materials, such as a ferromagnet and a non-magnetic material, is further limited by the spin-mixing conductance at the interface [7]. The rapidly diminishing spin current has severely hampered its exploitation. It is highly desirable to explore ways to enhance pure spin current.

Pure spin current phenomena and devices have employed ferromagnetic (FM) metals [3-5,10], FM insulators [8,9], and normal metals (NM) [3,8-10], where the FM magnetization sets the spin index of the spin current injected from the FM material, light NM (e.g., Cu) and heavy NM (e.g., Pt) respectively transmits and detects the spin current. Very recently spin current exploration involves antiferromagnetic (AF) materials [11-18]. The employment of antiferromagnets in spintronic devices is particularly attractive for terahertz (THz) devices [19]. Recently, spin pumping experiment in Pt/YIG (where YIG = $Y_3Fe_5O_{12}$) shows enhanced spin transport through an intervening AF NiO layer between YIG and Pt at room temperature [13,14]. It was suggested that the spin transport through the AF insulators is related to AF magnons and spin fluctuations [13,14], where the AF spins, strongly coupled to the precessing YIG magnetization, transport the spin current [13,14]. However, thus far, spin transport through AF insulators has only employed ferromagnetic resonance measurements (FMR) at the GHz frequency range [11,13-15,18], which is far less than the characteristic frequencies (up to 1 THz) of the AF NiO. The excitation and transmission of spin current, including amplification, through AF is far from clear. Coherent Néel dynamics employed to explain the spin transport and enhancement in such systems at room temperature [16], implies more prevalent spin transport enhancement at $T \ll T_N$. With the absence of the key experimental results, the mechanism for the large spin current enhancement observed at room temperature remains elusive [14]. The spin current amplification phenomena have thus far been observed in Pt/NiO/YIG and only at FMR frequencies. To unlock the underlying physics, it is essential to employ different spin current injection method, different AF materials, and a variety of metals other than Pt, and perform measurements over a wide temperature range. The comprehensive experimental studies would constrain the theory that accounts for the results.

In this Letter, we report enhanced spin current through AF (AF = NiO and CoO) generated by the longitudinal spin Seebeck effect (LSSE) in the layer structure of NM/AF/YIG over a wide temperature range. The pure spin current injected from YIG, transporting through the AF layer, is detected by the ISHE in various *3d, 4d,* and *5d* NM. In contrast to spin pumping, LSSE is a DC injection method without coherent resonance excitations at high frequencies. We show that the transmitted spin current detected in the NM has a maximum near the $T_N$ of the AF layer of a specific thickness, indicating the dominant roles of magnons and spin fluctuation in the AF on the spin transport, rather than the collective AF ordering dynamics. Equally important, we also demonstrate in various NMs that the spin current enhancement scales linearly with the spin-mixing conductance at the NM/YIG interface. Theoretical calculations of spin current enhancement and spin mixing conductance in such layer geometry are qualitatively consistent with the experimental results.

NiO is a well-known AF insulator with a face-centered cubic rock salt structure and a bulk Néel temperature of $T_N = 525$ K [20]. We used magnetron sputtering to fabricate polycrystalline multilayers onto polished polycrystalline YIG substrates 0.5 mm thick via

DC Ar sputtering for the NMs, reactive (Ar + O$_2$) sputtering for NiO and RF Ar sputtering for CoO at ambient temperature. The samples are denoted as Pt(3)/NiO(1)/YIG, where the numbers in parentheses are thickness in nm. The lateral sizes of all the rectangular samples are 7 mm × 2 mm. As shown in Fig. 1(a), the sample is thermally linked to a copper holder as a heat sink with its temperature measured by a thermocouple, while a heater is at the top of the sample surface with its temperature $T_{NM}$ measured via its electrical resistance. Between the heater and the heat sink, we established an out-of-plane temperature gradient $\nabla T$, for which most of temperature drop occurs in the thicker YIG and injects a pure spin current $J_S$ into the multilayer. The direction of $\nabla T$ dictates that of $J_S$. A small magnetic field aligns the YIG magnetization along the short direction of the sample that sets the spin index $\sigma$ of the pure spin current. The ISHE in the NM generates an electric field in the direction of $\sigma \times J_S$ with a voltage $V_{ISHE}$ along the long direction of the sample. In this open circuit DC measurement there is no high frequency coherent excitations. The applied magnetic field, less than 100 mT in magnitude, only aligns the YIG magnetization and does not alter appreciably the AF ordering in NiO.

The measured ISHE voltage $V_{ISHE}$ in NM/YIG and NM/NiO/YIG are shown in Fig. 1(b) and 1(c) as a function of the applied magnetic field. All the results have been obtained at $T_{NM}$ = 303 K in samples of the same size with and same $\nabla T$ = 10 K/mm. In Fig. 1(b), the results of Pt(3)/YIG (red curve), are similar to those previously observed by spin pumping [8,9]. With 1 nm thick NiO inserted between Pt(3) and YIG, $V_{ISHE}$ of Pt(3)/Ni(1)/YIG (blue curve) dramatically increases. The null results of $V_{ISHE}$ in the Pt/NiO/Si (black curve) shows that NiO itself does not generate any spin current at all.

The large enhancement of $V_{ISHE}$ due to the insertion of NiO also occurs in Ta/NiO/YIG (Fig. 1(c)), but that the polarity of the enhanced $V_{ISHE}$ in YIG/NiO/Ta is reversed due to its spin Hall angle of the opposite sign. In both cases, the inserted NiO layer greatly increases $V_{ISHE}$ while preserving the spin index. The enhancement of pure spin current due to the presence of the thin NiO spacer layer is clearly established.

Since $V_{ISHE}$ is proportional to the separation $L$ of the voltage leads and the temperature gradient $\Delta T/t_{YIG}$ within the YIG thickness $t_{YIG}$, we use the normalized parameter $S = \frac{V_{ISHE}/L}{\Delta T/t_{YIG}}$, also known as the transverse thermopower, that allows comparison of results taken under different experimental conditions. We use the ratio $S(t_{NiO})/S(0)$, where $S(t_{NiO})$ with, and $S(0)$ without, the presence of the NiO layer of thickness $t_{NiO}$. As shown in Fig. 2(a), both Pt/NiO/YIG and Ta/NiO/YIG, $S(t_{NiO})/S(0)$ increases sharply from 1, reaching a maximum at $t_{NiO} \approx 1$ nm before decreasing exponentially as shown in the inset of Fig. 2(a). The maximal value of $S(t_{NiO})/S(0)$ for Ta/NiO/YIG at $t_{NiO} \approx 1$ nm is higher, but its decay length $\lambda(Ta)_{NiO} = 1.3$ nm is considerably smaller than $\lambda(Pt)_{NiO} = 2.5$ nm for Pt/NiO/YIG. Similar behavior has also been observed in another AF insulator of CoO inserted between YIG and NM. Figure 2(b) shows in Ta/CoO/YIG, $S(t_{CoO})/S(0)$ has a maximum at $t_{CoO} \approx 2$ nm. With the CoO results, we show that the spin current enhancement phenomenon is not exclusive to NiO. To illustrate the unique feature of the intervening AF layer, we have also inserted $AlO_x$ between YIG and Pt. As shown in Fig. 2(c), $S(t_{AlOx})/S(0)$ in Pt(3)/$AlO_x$($t_{AlOx}$)/YIG at $T_{Pt} = 303$ K exhibits the expected exponential decay, monotonically decreasing with a very short decay length of $\lambda(Pt)_{AlOx} = 0.23$ nm, without enhancement at all.

In Fig. 3(a), we show the temperature dependences of the $S$ value from about 10 K to room temperature in Pt(3)/NiO($t_{NiO}$)/YIG for $t_{NiO} = 0$, 0.6, 1.2 and 2 nm, highlighting the strong temperature dependence and sensitivity to the NiO layer thickness. Without the NiO layer, the $S$ value of Pt(3)/YIG (labeled as 0 nm) is small, hardly varying for $T_{Pt}$ between 65 K and 300 K. However, with the insertion of the NiO layer, the large $S$ value of Pt/NiO/YIG acquires a very different temperature dependence exhibiting a well-defined broad peak. For $t_{NiO} = 0.6$, 1.2 and 2 nm, the peak temperature progressively increases, whereas the peak height changes sharply and non-monotonically from 4.3 µV/K to 6 µV/K and to 1.2 µV/K respectively.

The spin current injected into the NM layer is $J_S = \frac{t_{NM}\sigma_{NM}}{\theta_{SH}\lambda_{NM} tanh\left(\frac{t_{NM}}{2\lambda_{NM}}\right)} \frac{V_{ISHE}}{L}$, where $\sigma_{NM}$, $t_{NM}$, $\Theta_{SH}$, $\lambda_{NM}$ are the conductivity, the thickness, the spin Hall angle, and the spin diffusion length of the NM layer, respectively [21]. Then, we have $J_S = \frac{t_{NM}\sigma_{NM}}{\Theta_{SH}\lambda_N tanh\left(\frac{t_{NM}}{2\lambda_{NM}}\right)}\frac{S\Delta T}{t_{YIG}}$. Since the injected pure spin current $J_S$ in NM is proportional to the parameter $S(t_{NiO})$, the ratio $S(t_{NiO})/S(0)$ gives $J_S(t_{NiO})/J_S(0)$ for a temperature gradient in YIG, the amplification of pure spin current due to the presence of NiO. The results in Fig. 3(b) of $J_S(t_{NiO})/J_S(0)$ for Pt(3)/NiO($t_{NiO}$)/YIG appears similar to those shown in Fig. 3(a), because $S(0)$ for Pt/YIG without NiO varies little except at low temperatures. As shown in Fig. 3(b), the presence of the intervening NiO layer greatly enhances the spin current, up to a factor of 11.6 for the 1.2 nm thick NiO.

The enhancement of $J_S$ has a well-defined peak at $T_{peak}$, whose values of 142 K, 191 K and 263 K depend strongly with $t_{NiO}$ = 0.6 nm, 1.2 nm and 2 nm, respectively. As shown in Fig. 3(c), the $T_{peak}$ increases linearly with the $t_{NiO}$ as $t_{NiO}$ < 2 nm. Similar behavior has been also observed recently in IrMn/Cu/NiFe by spin pumping [18]. The $T_{peak}$ is near the reduced intrinsic Néel temperature $T_N(t_{NiO})$ of the isolated thin NiO layer due to finite size effects [22,23]. The value of $T_N(t_{NiO})$ of NiO thin film can be estimated by the blocking temperature at which exchange bias of a ferromagnetic layer exchange-coupled to the NiO vanishes [24]. The blocking temperature is close to and usually slightly lower than $T_N$ [22]. As shown in the inset of Fig. 3(d), the magnetic hysteresis loop of a NiO(1)/Co(3) film shifts to negative field at $T$ = 90 K after cooling from 330 K under a 0.5 T field, due to the exchange bias [24,25]. From the temperature dependence of the exchange bias field shown in Fig. 3(d), the blocking temperature of 1 nm thick NiO layer is around 170 K, which agrees with Ref. 25. The $S$ values in all cases, with or without NiO, as shown in Fig. 3(a), decreases towards zero as $T_{Pt}$ approaches 0 K due to the lack of thermal excitations of magnons in YIG at low temperatures [26,27].

To address the physics of the observed behavior, we calculated the spin current transmission under an out-of-plane temperature gradient in NM/AF/F. In contrast to coherent zero-wave number magnons for spin pumping, the spatial dependent non-equilibrium thermal magnons have a broad spectrum distribution [27], and thus it is possible to transfer one F magnon to one AF magnon via interface exchange interaction. The spin currents in the NM for NM/F and NM/AF/F can be expressed as

$$J_S^{\text{NM/F}} = \kappa \nabla T \frac{G_{\text{NM}} G_{\text{NM/F}}}{G_{\text{NM}} G_{\text{F}} + (G_{\text{F}} + G_{\text{NM}}) G_{\text{NM/F}}} e^{-\frac{x}{\lambda_{\text{NM}}}} \quad (1)$$

$$J_S^{\text{NM/AF/F}} = \kappa \nabla T \frac{1}{\cosh\left(\frac{t_{\text{AF}}}{\lambda_{\text{AF}}}\right) + \delta \sinh\left(\frac{t_{\text{AF}}}{\lambda_{\text{AF}}}\right)} \frac{G_{\text{NM}} G_{\text{NM/AF}}}{G_{\text{NM}} G_{\text{F}} + (G_{\text{F}} + G_{\text{NM}}) G_{\text{NM/AF}}} e^{-\frac{x}{\lambda_{\text{NM}}}}, \quad (2)$$

where $\kappa$ is the spin current coefficient due to the temperature gradient, $G_{\text{NM}}$, $G_{\text{F}}$, $G_{\text{AF}}$, $G_{\text{NM/F}}$ and $G_{\text{NM/AF}}$ are the spin current conductance of bulk NM, bulk F, bulk AF, the NM/F interface, the AF/F interface and the NM/AF interface, respectively. $\lambda_{\text{NM}}$ is the spin diffusion length of the NM, $\lambda_{\text{AF}}$ the magnon decay length of the AF, $t_{\text{AF}}$ the AF thickness, and $\delta = G_{\text{AF}}\left(\frac{1}{G_{\text{F}}} + \frac{1}{G_{\text{AF/F}}}\right)$. Then, the spin current ratio is

$$\frac{J_S^{\text{NM/AF/F}}}{J_S^{\text{NM/F}}} = \left[1 + \frac{(a-1)G_{\text{NM}}}{G_{\text{NM/AF}} + G_{\text{NM}}}\right] \frac{1}{\cosh\left(\frac{t_{\text{AF}}}{\lambda_{\text{AF}}}\right) + \delta \sinh\left(\frac{t_{\text{AF}}}{\lambda_{\text{AF}}}\right)}, \quad (3)$$

where $a = \frac{G_{\text{NM/AF}}}{G_{\text{NM/F}}}$. In the spin wave approximation, $G_{\text{NM/AF}}$ scales as $\left(J_{sd}^{\text{AF}}\right)^2 \left(\frac{T}{T_N}\right)^4$ and $G_{\text{NM/F}}$ scales as $\left(J_{sd}^{\text{F}}\right)^2 \left(\frac{T}{T_C}\right)^{3/2}$ [27], where $J_{sd}^{\text{AF}}$ and $J_{sd}^{\text{F}}$ are exchange constants of the AF and the F, respectively, and $T_C$ Curie temperature of the F. Then, $\frac{G_{\text{NM/AF}}}{G_{\text{NM/F}}} = b\left(\frac{J_{sd}^{\text{AF}}}{J_{sd}^{\text{F}}}\right)^2 \left(\frac{T}{T_N}\right)^4 \left(\frac{T}{T_C}\right)^{-3/2}$ ($b$ is a numerical constant of order of 1). The $T_N$ of NiO is lower than the $T_C$ of YIG. Thus, $G_{\text{NM/AF}}$ increases much faster with $T$. At the high temperature $G_{\text{NM/AF}}$ is larger than $G_{\text{NM/F}}$. This is primary due to enhanced AF magnons or enhanced spin fluctuation in NiO. Thus, the significant enhancement of spin current occurs near $T_N$ in agreement with experiments.

The enhancement of spin current decreases but still pronounced in a large temperature range above $T_N$ in the absence of long range AF ordering. This indicates the prominent roles of spin fluctuation and short range spin correlation in AF on spin transport [17,18,28]. Note that short range spin correlation in AF still exists at temperatures much higher than $T_\text{N}$, as revealed by neutron scattering [29]. Above the $T_\text{N}$, the magnons whose wavelength shorter than the spin correlation length remains. The spin correlation length of NiO is $\xi = l\left(\frac{T}{T_\text{N}} - 1\right)^{-0.64}$, where $l$ = 0.55 nm [29]. The number of magnons participating the spin transport decreases due to the loss of the magnons whose wavelength longer than the spin correlation length. Thus, the spin current enhancement is maximum near the $T_\text{N}$. For thicker NiO layers with $t_\text{NiO}$ > 3.5 nm, there is no appreciable enhancement because of the drastic decay of spin current. Overall, the largest enhancement occurs near $t_\text{NiO} \approx 1$ nm, as shown in Fig. 2(a).

As discussed above, the observed enhancement of spin transport in the NM/AF/YIG is attributed to the large spin conductance at both the NM/AF and the AF/YIG interfaces. We experimentally explore this essential feature in spin current enhancement in NM/NiO/YIG with various NMs in addition to Pt, the only metal studied to date. We determine $J_\text{S}(t_\text{NiO})/J_\text{S}(0)$ with $t_\text{NiO}$ = 1 nm at $T_\text{NM}$ = 303 K for *3d* (Cr, Mn), *4d* (Pd), and *5d* (Ta, W, Pt, Au) metals. The spin-mixing conductances at the NM/YIG interfaces have been measured from the FMR linewidth in spin pumping [30-33]. Figure 4(a) shows our measured values of $J_\text{S}(1)/J_\text{S}(0)$ for various NMs vs. the spin-mixing conductance $G^{\uparrow\downarrow}_\text{NM/YIG}$ reported by spin pumping at room temperature in NM/YIG [30-33]. The $G^{\uparrow\downarrow}_\text{NM/YIG}$ values in units of $10^{18}$ m$^{-2}$ in ascending order for NM = Cr, Pd, W, Au, Pt, Mn, Ta are 0.83, 1.1, 1.2, 2.7, 3.9, 4.5, and 5.4, respectively [30-33]. Most remarkably, $J_\text{S}(1)/J_\text{S}(0)$ is *linearly proportional* to $G^{\uparrow\downarrow}_\text{NM/YIG}$, i.e., $J_\text{S}(1)/J_\text{S}(0) = CG^{\uparrow\downarrow}_\text{NM/YIG}$, where $C$ = 8.5 × $10^{-19}$ m$^2$. In NM/YIG, the spin current transmission is dictated by the spin-mixing conductance at the NM/YIG. With the insertion of a NiO layer in NM/NiO/YIG, the spin fluctuation in the thin AF NiO layer *amplifies* the spin current transmission.

The ratio of spin current in the NM between the NM/F and the NM/AF/F can be calculated from Eq. (3). As shown in Fig. 4(b), the calculated spin current enhancement in NM/NiO(1)/YIG increases with the spin-mixing conductance in NM/YIG. The calculated spin current enhancement is consistent with the linear correlation observed experimentally at room temperature.

It may be noted that spin Hall angle $\Theta_{SH}$, an important property of the NM in converting pure spin current, does *not* play a role in spin current enhancement. In particular, while Cr, Ta, W and Pt have large $\Theta_{SH}$ values [30,32], only Ta and Pt have large $J_S(1)/J_S(0)$ above 3, whereas those of Cr and W have small values of less than 1, i.e., only reduction. The linear behavior of $J_S(1)/J_S(0)$ with $G_{NM/YIG}^{\uparrow\downarrow}$, provides an essential criterion for selecting materials for large spin current enhancement. We also note that such spin current enhancement is observed with 1 nm thick paramagnetic NiO at room temperature (above $T_N$), thus not related to coherent AF ordering dynamics.

In conclusion, we have observed spin current enhancement through AF by DC thermal injection in a broad temperature range in various metals. The spin conductance can be enhanced in NM/AF/YIG due to the magnons and spin fluctuation in the thin AF layer. The degree of enhancement increases with the spin-mixing conductance at the NM/YIG interface. These key results provide the criteria for selecting materials with effective spin current enhancement.

*Note added*   Recently, Ref. 34 accounts for the AF insulator thickness dependence of spin current in NM/AF/F by diffusive thermal AF magnons. Ref. 35 describes the spin transport through magnetic insulator below and above the magnetic transition temperature by Heisenberg interactions using auxiliary particle methods.

This work was supported by the U.S. Department of Energy, Office of Science, Basic Energy Science, under Award Grant No. DE-SC0009390. W. L. was supported in part by C-SPIN, one of six centers of STARnet, a SRC program sponsored by MARCO and DARPA. K. C. and S. Z. were supported by National Science Foundation under Grant number ECCS-1404542. W. L. thanks Ssu-Yen Huang from National Taiwan University for fruitful discussions.

**References**

[1] J. E. Hirsch, Phys. Rev. Lett. **83**, 1834 (1999).

[2] S. Zhang, Phys. Rev. Lett. **85**, 393 (2000).

[3] T. Kimura *et al.*, Phys. Rev. Lett. **98**, 156601 (2007).

[4] M. Johnson, Phys. Rev. Lett. **70**, 2142 (1993).

[5] F. J. Jedema, A. T. Filip, and B. J. van Wees, Nature **410**, 345 (2001).

[6] R. Urban, G. Woltersdorf, and B. Heinrich, Phys. Rev. Lett. **87**, 217204 (2001).

[7] Y. Tserkovnyak, A. Brataas, and G. E.W. Bauer, Phys. Rev. Lett. **88**, 117601 (2002).

[8] K. Uchida *et al.*, Appl. Phys. Lett. **97**, 172505 (2010).

[9] S. Y. Huang, *et al.*, Phys. Rev. Lett. **109**, 107204 (2012).

[10] E. Saitoh, M. Ueda, H. Miyajima, and G. Tatara, Appl. Phys. Lett. **88**, 182509 (2006).

[11] C. Hahn *et al.*, EPL **108**, 57005 (2014).

[12] J. B. S. Mendes *et al.*, Phys. Rev. B **89**, 140406(R) (2014).

[13] H. Wang, C. Du, P. Ch. Hammel, and F. Yang, Phys. Rev. Lett. **113**, 097202 (2014).

[14] H. Wang, C. Du, P. Ch. Hammel, and F. Yang, Phys. Rev. B **91**, 220410(R) (2015).

[15] T. Moriyama *et al.*, Appl. Phys. Lett. **106**, 162406 (2015).

[16] S. Takei, T. Moriyama, T. Ono, and Y. Tserkovnyak, Phys. Rev. B **92**, 020409(R) (2015).

[17] Z. Qiu *et al.*, arXiv:1505.03926v2.

[18] L. Frangou *et al.*, Phys. Rev. Lett. **116**, 077203 (2016).

[19] T. Kampfrath *et al.*, Nat. Photonics **5**, 31 (2011).

[20] P. W. Palmberg, R. E. DeWames, and L. A. Vredevoe, Phys. Rev. Lett. **21**, 682 (1968).

[21] C. Du *et al.*, Phys. Rev. Lett. **111**, 247202 (2013).

[22] T. Ambrose and C. L. Chien, Phys. Rev. Lett. **76**, 1743 (1996).

[23] D. Alders *et al.*, Phys. Rev. B **57**, 11623 (1998).

[24] T. Ambrose and C. L. Chien, J. Appl. Phys. **83**, 6822 (1998).

[25] Z. Y. Liu and S. Adenwalla, J. Appl. Phys. **94**, 1105 (2003).

[26] J. Xiao *et al.*, Phys. Rev. B **81**, 214418 (2010).

[27] S. S.-L. Zhang and S. Zhang, Phys. Rev. B **86**, 214424 (2012).

[28] Y. Ohnuma, H. Adachi, E. Saitoh, and S. Maekawa, Phys. Rev. B **89**, 174417 (2014).

[29] T. Chatterji, G. J. McIntyre, and P.-A. Lindgard, Phys. Rev. B **79**, 172403 (2009).

[30] H. Wang *et al.*, Phys. Rev. Lett. **112**, 197201 (2014).

[31] C. Du, H. Wang, F. Yang, and P. Ch. Hammel, Phys. Rev. Appl. **1**, 044004 (2014).

[32] C. Du, H. Wang, F. Yang, and P. Ch. Hammel, Phys. Rev. B **90**, 140407(R) (2014).

[33] L. Ma *et al.*, arXiv:1508.00352v3.

[34] S. M. Rezende, R. L. Rodríguez-Suárez, and A. Azevedo, Phys. Rev. B **93**, 054412 (2016).

[35] S. Okamoto, Phys. Rev. B **93**, 064421 (2016).

**Figure Captions**

FIG. 1  (a) Schematic of thermal spin transport measurement. Inverse spin Hall voltage $V$ as a function of the applied field $H$ in (b) Pt(3)/YIG, Pt(3)/NiO(1)/YIG and Pt(3)/NiO(1)/SiO$_x$/Si, (c) Ta(3)/YIG and Ta(3)/NiO(1)/YIG. The temperature of the metal layer is about 303 K, and the out-of-plane temperature gradient cross the YIG is about 10 K/mm. The number in the layered structure denotes thickness in nm.

FIG. 2  (a) Transverse thermopower $S(t_{NiO})$ normalized by $S(0)$, the transverse thermopower without NiO, of Pt(3)/NiO($t_{NiO}$)/YIG and Ta(3)/NiO($t_{NiO}$)/YIG as a function of the NiO thickness $t_{NiO}$ at $T_{NM}$ = 303 K. Inset shows $S(t_{NiO})/S(0)$ in the logarithmic scale as a function of $t_{NiO}$. (b) $S(t_{CoO})/S(0)$ as a function of CoO thickness $t_{CoO}$ in Ta(3)/CoO($t_{CoO}$)/YIG at $T_{Ta}$ = 303 K. (c) $S(t_{AlOx})/S(0)$ as a function of AlO$_x$ thickness $t_{AlOx}$ in Pt(3)/AlO$_x$($t_{AlOx}$)/YIG at $T_{Pt}$ = 303 K. Inset shows $S(t_{AlOx})/S(0)$ in the logarithmic scale as a function of $t_{AlOx}$.

FIG. 3  Temperature dependences of (a) $S$ and (b) $J_S(t_{NiO})/J_S(0)$ in Pt(3)/NiO($t_{NiO}$)/YIG for various NiO thicknesses $t_{NiO}$. In (b), $J_S(t_{NiO})/J_S(0)$ for $t_{NiO} \neq 0$ has a peak at the peak temperature $T_{peak}$, whereas the dashed line denotes $J_S(0)/J_S(0) = 1$. (c) Peak temperature $T_{peak}$ as a function of $t_{NiO}$. (d) Temperature dependence of exchange bias field in a NiO(1)/Co(3) film. Inset shows the magnetic hysteresis loop of the NiO(1)/Co(3) film at $T$ = 90 K after the field cooling.

FIG. 4  (a) Our measured spin current enhancement in NM(3)/NiO(1)/YIG for various NM at $T_{NM}$ = 303 K vs. the measured spin-mixing conductance in NM/YIG (from Refs. 30-33). (b) Calculated spin current enhancement in NM/NiO(1)/YIG as a function of the spin mixing conductance at the NM/YIG interface, in the case of $J_{sd}^{NiO} = 2J_{sd}^{YIG}$, $\lambda_{NiO}$ = 2.5 nm, $T_C$ = 560 K, $T_N$ = 190 K and $T$ = 300 K.

**Figures**

FIG. 1

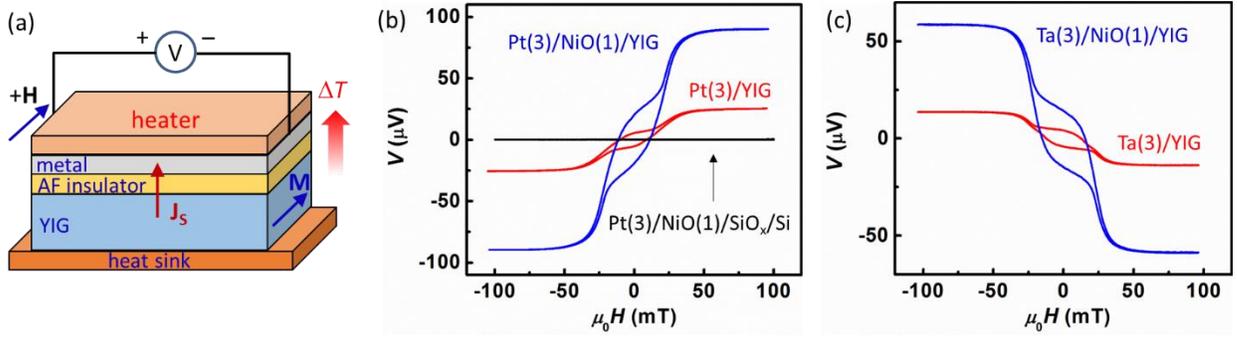

FIG. 2

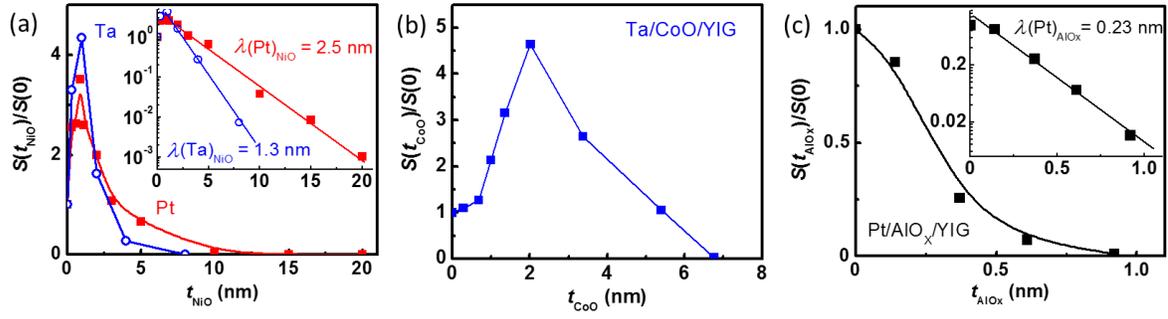

FIG. 3

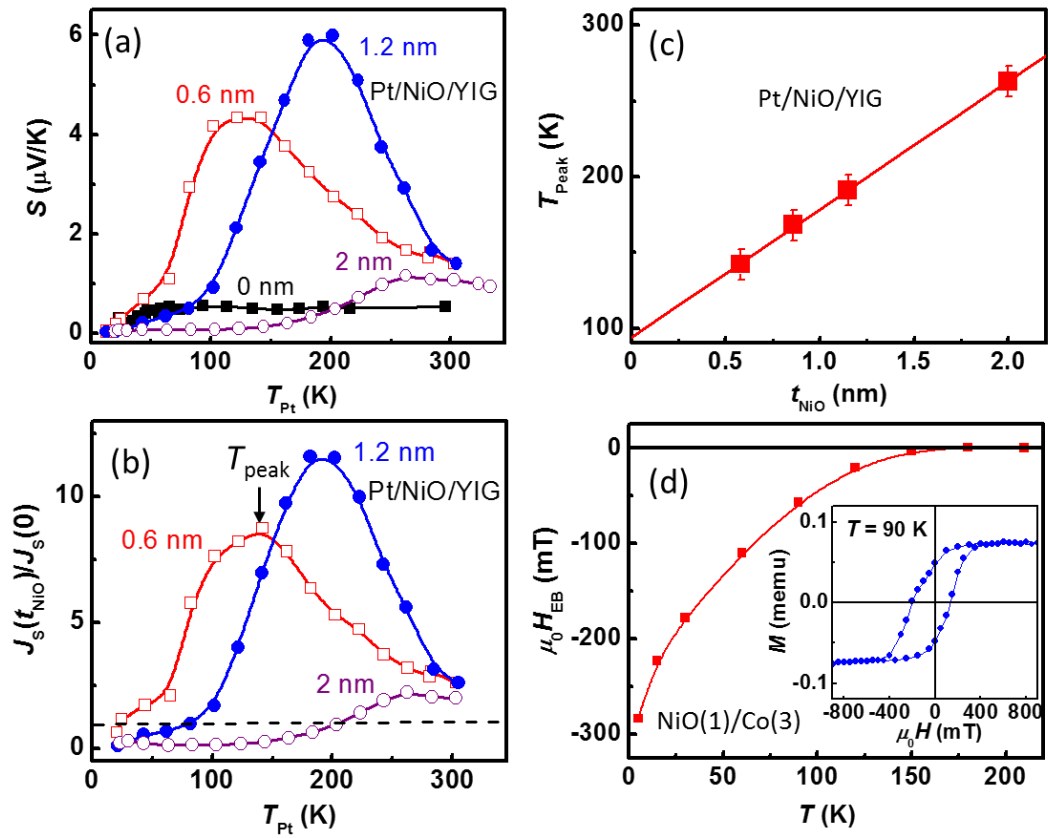

FIG. 4

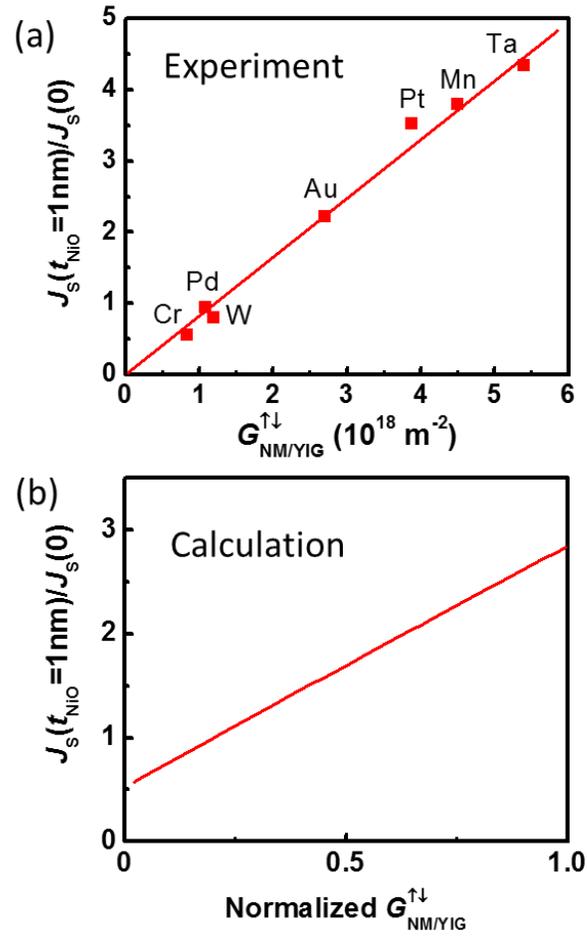